# Microrheology of non mulberry silk varieties by optical tweezer and video microscopy based techniques

Yogesha[1], Raghu A [1,2], Siddaraju G N [3], G Subramanya[4], Somashekar R[3], and Sharath Ananthamurthy [1,*]

[1]Department of Physics, Jnanabharathi Campus, Bangalore University, Bengaluru-560056, India
[2]Department of Physics, Government College for Women, Mandya-571401, India
[3]Deparment of Studies in Physics, University of Mysore, Mysore-570006, India
[4]Department of Studies in Sericulture, University of Mysore, Mysore-570006, India,

*Email:asharath@gmail.com

**ABSTRACT**
We have carried out a comparative study of the microrheological properties of silk fibroin solutions formed from a variety of silks indigenous to the Indian subcontinent. We present the measured viscoelastic moduli of Tasar silk fibroin solution using both a single and dual optical tweezer at 0.16% and 0.25% (w/v). The bandwidth of the measurements carried out using optical tweezers is extended down to the lower frequency regime by a video microscopy measurement. Further, we have measured the viscoelastic moduli of Eri and Muga varieties of silk fibroin solutions at a higher concentration (1.00% w/v) limiting the tool of measurement to video microscopy, as the reduced optical transparencies of these solutions at higher concentration preclude an optical tweezer based investigation. The choice of a higher concentration of fibroin solution of the latter silk varieties is so as to enable a comparison of the shear moduli obtained from optical methods with their corresponding fibre stiffness obtained from wide angle X-ray scattering data. We report a correlation between the microstructure and microrheological parameters of these silk varieties for the concentration of fibroin solutions studied.

Keywords: Microrheology, silk fibres, optical tweezer, video microscopy, WAXS.

## INTRODUCTION

Silk is a fibrous protein polymer, secreted as a continuous filament by silk worms, and spiders. Silk worms and spiders manufacture silk and use it to engineer structures such as cocoons, webs and nets. It has attracted the interests of scientists and engineers for varied reasons from a long time. The initial interest was in its use in manufacturing textiles of high strength, durability and aesthetic value. Silk has been found to be a valuable biotechnological material because of its compatibility with blood and its high permeability to both water and oxygen [1]. In order for use in a specific biomedical application, it should be regenerated in a desirable form. By dissolving the silk fibre in a suitable solvent, a Regenerated Silk Fibroin (RSF) solution can be prepared.

There are two classes of silks namely, mulberry (*Bombyx mori*) and non mulberry varieties. India is the only country known for the production of four varieties of silk. In the literature, a number of studies have been reported on the structure and mechanical properties of mulberry as well as non mulberry silks, including the tensile stress-strain and recovery behavior of Indian silk fibres under stress [2] revealing properties such as its high breaking extension and strength. The microstructure of the Indian silk varieties has been elucidated by studying their amino acid contents [3]. Stress relaxation in non-mulberry silks has been observed when compared to that in mulberry silk [4]. Studies on the solubility and rheological behavior of silk fibroin (*Bombyx mori*) using a rheometer have been reported [1].

Rheological characterization of soft materials in bulk yields the viscoelastic parameters of the sample averaged over the volume of the sample studied. The advantage offered by microrheology is that much smaller sample volumes are needed for characterization, a matter of significance when the sample is scarce or expensive to obtain in larger amounts. Local structural heterogeneity found in entangled polymer networks such as silk fibroin solutions, can result in a variation in the viscoelastic properties from region to region, features that bulk measurements may fail to capture [5].

In microrheology, the position information of the tracer particles embedded in a complex fluid are recorded. The motion of the beads is then interpreted in terms of the viscoelastic properties of the surrounding medium. Since each tracer particle reflects the local mechanical response of the surrounding material, microrheology can yield this information on micro and even nano length scales in soft matter.

A single trapped bead in an optical tweezer enables one-point microrheology. This is a sensitive technique to characterize the local environment within the sample at micro and nano length scales. However, in the absence of heterogeneities, or where one is more concerned with measuring the bulk limit of the viscoelastic parameters of the material, differences in the microrheological parameters arising due to local variations in structures need to be averaged. In such situations one can monitor the cross correlated fluctuations of a pair of beads, whose relative positions varies with time [6]. This is achieved with a dual optical tweezer [7,8].

We present a single and dual optical tweezer based study of the microrheology of Tasar (Karnataka, India) silk fibroin solution, at a temperature of 22±0.5$^{o}$C for two concentrations (w/v). We extend the bandwidth of measurements to lower frequencies by including a video microscopic analysis. This study enables a comparative assessment of the single and dual optical tweezer based methods in obtaining microrheological parameters.

We have carried out measurements of the frequency dependent shear moduli of non mulberry silk fibroin solutions formed from Eri[†] and Muga[‡] (Assam, India) at a higher concentration(1.00% w/v) than that for Tasar[§]. The limited transparencies of the non mulberry silk fibroin solutions, at these concentrations preclude optical tweezer based studies and hence, for these varieties we report only the video microscopy based microrheological results. We have chosen a higher concentration of the latter silk varieties in order to compare the measured shear moduli, with their corresponding single fibre stiffness obtained from a Wide Angle X-ray Scattering (WAXS) analysis. It is reasonable to expect the fibre stiffness to have a bearing, if any, on the shear modulus of the corresponding silk fibroin solution at a higher concentration and hence the choice of 1.00% w/v. From our measurements we record that silk fibres of Muga forming the fibroin solution shows higher shear modulus and also possesses greater fibre stiffness, than the corresponding properties of Eri variety.

---

[†] Eri silk comes from the worm *Samia cynthia ricini*, found in North East of India and some parts of China and Japan. Eri silkworm secretes white or brick-red silk. Eri silk is not a continuous filament unlike other silks, and is thicker and heavier than other silks.

[‡] Muga silk comes from the worm *Antheraea Assama*, is widely distributed and cultured in North-Eastern India particularly in the state of Assam, India. It is popular for its natural golden colour, glossy fine texture and durability.

[§]Tasar Silk comes from the worm *Antheraea mylitta*, and is a natural fauna of tropical India. It is white in colour.

This paper is organized as follows: After a brief discussion of the methods for preparing the fibroin solution samples of all varieties of silk fibres we present the optical tweezer based microrheological measurements using single and dual bead setups. This is followed by video microscopic analysis to extract the viscoelastic data. Finally, we present results obtained using all the methods and provide a comparative study with micro-structural data of the fibres gathered through WAXS.

**SAMPLE PREPARATION**

A 3:1(weight ratio) mixture of calcium nitrate tetra hydrate (Ca $(NO_3)_2$ $4H_2O$) and absolute methanol ($CH_3OH$) is prepared and is used as a solvent for non mulberry silk fibres [9]. Though this recipe has been developed for mulberry (*Bombyx mori*) silk, the same recipe is found to work for non mulberry silks. Raw silk is first degummed thoroughly, as reported elsewhere [9,10]. The degummed silk is cut into small pieces and then added to the solvent. The mixture is stirred well at a temperature of 70°C, until the silk fibre dissolves completely in the solvent to obtain undialyzed RSF. The solutions of a given concentration are prepared by dissolving a known weight of silk in a definite volume of solvent. The RSF solution at two different concentrations viz., 0.16% and 0.25% (w/v) are prepared by the above methods for Tasar silk, and 1.0% for both Eri and Muga silk.

Polystyrene beads (Cat. no. 17134, 2.799 µm, Polysciences Inc., USA) are mixed with RSF solution in low concentrations (2µL in 1mL of RSF) to avoid inter-bead interactions and they are used as probes to measure the stress-deformation relation in the material. The sample was taken in a well constructed by a rubber 'O' ring (10 mm diameter & 2 mm thickness) on a cover slip and is left for 20 minutes to stabilize before taking the measurements. For each trial, around 30 µL of sample was used and the temperature was maintained at 22±0.5 °C. The chemicals and reagents used in this work were HPLC grade supplied by Merck Chemicals, India.

**MEASUREMENT TECHNIQUES**

A detailed description of the optical tweezer with image processing setup used in this work can be found elsewhere [11]. We present here the method of calibrating the imaging system and its use in obtaining microrheological parameters.

**Optical tweezer based technique**

In the optical tweezer technique, for 1-point microrheology, a bead of radius '$a$' embedded in a given sample is trapped with a laser and the bead's position is tracked either by the same laser or with a different laser back scattered from the trapped bead. A high speed quadrant photo detector (EOS, USA) interfaced with a high speed DAQ card (PCI 6143, NI, USA), is used to acquire the bead's position with a scan rate of $10^5$ samples/sec for five seconds. 20 such data sets are stored on the hard drive of a PC, for a given trapped bead, at similar temperatures and surroundings. In 2 point microrheology, two beads, each of radii '$a$' that are separated by a distance of '$r$' are trapped by two laser beams (Figure 1.) with different planes of polarization. The laser beams backscattered by the two beads, therefore can be separated with the use of a polarizing beam splitter. The two bead's positions are tracked simultaneously with two quadrant detectors at a scan rate of $10^5$ samples/sec for five seconds using a simultaneous 8 channel DAQ card (PCI 6143, NI, USA). From these data, the cross correlated and auto-correlated PSD is calculated using the relation (3). Since the PSD obtained from 1-point as well as 2-point techniques, contains the effect of the inherent trap stiffness, the data needs to be corrected for this trap stiffness, to obtain the unbiased microrheological parameters. The method followed for this correction is explained below.

A detailed analysis procedure for 2-point microrheology can be found in [12]. We present a summary of the relevant theory here.
The complex response function $\chi_{\alpha\beta}^{j,k}(\omega) = \chi_{\alpha\beta}^{jk}{}'(\omega) + i\chi_{\alpha\beta}^{jk}{}''(\omega)$ of a bead can be related to the Fourier transform of the displacement $x_\alpha^j(\omega)$ of a particle j (1 or 2) in the direction $\alpha$ (x or y) and the Fourier transform of the applied force $F_\beta^k(\omega)$ to the particle k in the direction β by an equation:

$$x_\alpha^j(\omega) = \chi_{\alpha\beta}^{j,k}(\omega) F_\beta^k(\omega) \tag{1}$$

In thermal equilibrium and in the absence of external forces the fluctuation-dissipation theorem relates the imaginary part of the (single or inter-particle) response function to the equilibrium fluctuation spectrum of $x_\alpha^j(\omega)$:

$$\chi_{\alpha\beta}^{''(j,k)}(\omega) = \frac{\omega}{2k_B T} S_{\alpha\beta}^{(j,k)}(\omega) \tag{2}$$

Where $k_B T$ is the thermal energy and $S_{\alpha\beta}^{(j,k)}(\omega)$ is the single sided PSD given by,

$$S_{\alpha\beta}^{(j,k)}(\omega) = \int \langle u_\alpha^{(j)}(t) u_\beta^{(k)}(0) \rangle e^{i\omega t} dt \tag{3}$$

If we choose a coordinate system with the line connecting the centres of the two particles as along x axis and y spanning the plane perpendicular to the optical axis (laser propagation direction), all $S_{\alpha\beta}^{(j,k)}(\omega)$ are identically zero for $\alpha \neq \beta$. The remaining nonzero components of $S_{\alpha\beta}^{(j,k)}(\omega)$ are $S_{xx}^{(j,k)}(\omega)$ and $S_{yy}^{(j,k)}(\omega)$ for $\alpha = \beta$ in a dual trap, whereas in a single trap $S_x^{(j)}(\omega) = S_x^{(j,j)}(\omega)$ and $S_y^{(j)}(\omega) = S_y^{(j,j)}(\omega)$.

Once the imaginary part of the response function is calculated, the Kramers-Kronig relation can then be used to obtain the real part of the response function, provided that $\chi_{\alpha\beta}^{''(j,k)}(\omega)$ is known over a large enough frequency range, by an equation

$$\chi_{\alpha\beta}^{'(j,k)}(\omega) = \frac{2}{\pi} \int_0^\infty dt \cos(\omega t) \int_0^\infty d\xi\, \chi_{\alpha\beta}^{''(j,k)}(\zeta) \sin(2\pi\zeta t) \tag{4}$$

For an isolated particle and in the absence of the optical trap, the complex response function $\chi^{*(j)}(\omega)$ is related to the complex shear modulus $G$ of the inhomogeneous, isotropic and incompressible medium, by a generalized Stokes-Einstein relation [12,13],

$$\chi^{*(j)}(\omega) \to \alpha^{*(j)}(\omega) = \frac{1}{6\pi a G(\omega)} \tag{5}$$

where, $a$ is the radius of the particle and is the same for both particles in our experiments. $\alpha$ is the corrected response function from the measured $\chi$. It is noted that the measured response function $\chi$ directly reflects the actual rheology only in the absence of the trapping potentials.
Similarly, in the absence of traps, the inter-particle response functions $\chi_{xx,yy}$ are given by generalizations of the Oseen tensor [15]

$$\chi_{xx}(\omega) \to \alpha_{xx}(\omega) = \frac{1}{4\pi r G(\omega)} \quad \text{and}$$

$$\chi_{yy}(\omega) \to \alpha_{yy}(\omega) = \frac{1}{8\pi r G(\omega)} \tag{6}$$

Where, r is the distance between the two particles

$$\alpha_{xx} = \frac{\chi_{xx}}{1 - k^{(1)}\chi_{xx}^{(1)} - k^{(2)}\chi_{xx}^{(2)} - k^{(1)}k^{(2)}(\chi_{xx})^2 + k^{(1)}k^{(2)}\chi_{xx}^{(1)}\chi_{xx}^{(2)}}$$

$$\alpha_{yy} = \frac{\chi_{yy}}{1 - k^{(1)}\chi_{yy}^{(1)} - k^{(2)}\chi_{yy}^{(2)} - k^{(1)}k^{(2)}(\chi_{yy})^2 + k^{(1)}k^{(2)}\chi_{yy}^{(1)}\chi_{yy}^{(2)}} \tag{7}$$

**Video microscopy**

Movies containing nearly 1000 images of the freely diffusing, and thermally driven beads in a given medium are stored on to a PC, after focusing the microscope well above (about 20 µm) the cover slip wall [14]. Through multi particle tracking 2-5 beads, embedded in a medium, are tracked simultaneously for each field of view using image analysis algorithms (IDLVM, RSI Systems, USA) to obtain their time-sequenced positions [15,16]. The spatial resolution of our imaging system, determined by camera pixel density and optical magnification of the imaging system is 75nm/pixel at 100 fps, [14].

The position information is used to calculate the mean squared displacement (MSD=$\langle \Delta r^2(\tau) \rangle$) of beads in that medium at a temperature T, using a standard relation [17,18]. For each measurement, an ensemble averaged MSD of 25-30 beads is calculated. The accuracy of our measurements is tested by analyzing the ensemble averaged MSD of 30 stuck beads in water. For this, the water sample with beads is left for 2 days, so that beads sediments and stick to the cover slip. The stuck beads' motion is monitored for nearly one hour under the microscope, before taking position measurements. Then the stuck beads' are monitored for nearly 1000 frames and their ensemble averaged MSD ($\langle \Delta r^2(\tau) \rangle$) is calculated. This MSD is a measure of the static error '$\bar{\varepsilon}$' involved in the measurement, given by $\bar{\varepsilon} = \sqrt{\langle \Delta r^2(\tau) \rangle / 2}$, which is nearly 60 nm as shown in figure 2(a) [19]. This error is found to be almost constant at all lag times. The effect of inherent error of the system seems to be significant at the lower lag times and till about 0.2 s. Beyond 0.2 s, the MSD curve of stuck beads lies much below that of the free beads, thereby with insignificant effect on the actual MSD.

Static error free MSD $\langle \Delta r^2(\tau) \rangle_{sfree}$ is obtained by removing the error present in the measured MSD of free beads $\langle \Delta r^2(\tau) \rangle_{me}$ using the relation [19, 20]:

$$\langle \Delta r^2(\tau) \rangle_{sfree} = \langle \Delta r^2(\tau) \rangle_{me} - 2\bar{\varepsilon}^2 \tag{8}$$

MSD of stuck bead (static error) and free bead's MSD, before and after the correction of error is shown in figure 2(a). After the correction, the lag time dependent MSD has a slope of nearly 1 as expected for a pure viscous liquid. For a viscoelastic medium, MSD can be related to the complex shear modulus $G^*(\omega)$ through the generalized Stokes-Einstein relation (GSER) given by [21, 22]:

$$G^*(\omega) = \frac{k_B T}{\pi a \langle \Delta r^2(1/\omega) \rangle \Gamma[1+\alpha(\omega)] (1+\beta(\omega)/2)} \qquad (9)$$

where, '$k_B T$' is the Boltzmann constant and $\alpha(\omega)$ and $\beta(\omega)$ are the first and second order logarithmic time derivatives of the MSD. From the complex shear modulus $G^*(\omega)$ is obtained and hence the storage modulus $G'(\omega)$ and loss modulus $G''(\omega)$ of the material are found, by the analytical continuation method developed by Mason et al [21]

$$G'(\omega) = G^*(\omega) \{1/[1+\beta'(\omega)]\} \cos\left[\frac{\pi \alpha'(\omega)}{2} - \beta'(\omega)\alpha'(\omega)\left(\frac{\pi}{2}-1\right)\right]$$

$$G''(\omega) = G^*(\omega) \{1/[1+\beta'(\omega)]\} \sin\left[\frac{\pi \alpha'(\omega)}{2} - \beta'(\omega)[1-\alpha'(\omega)]\left(\frac{\pi}{2}-1\right)\right] \qquad (10)$$

where, $\alpha'(\omega)$ and $\beta'(\omega)$ are the first and second order logarithmic frequency derivatives of $G^*(\omega)$ and are obtained by fitting the data locally to a second order polynomial. All the microrheological parameters were obtained by using custom programs, written in LabVIEW.

**CALIBRATION OF THE SETUP**

For calibration, double distilled water - a pure viscous fluid is used, and measurements are carried out at a temperature of 22±0.5 °C for both the video microscopy and the optical tweezer work.

**Video microscopy:** Using the corrected MSD data, the frequency-dependent shear, storage and loss moduli are calculated. The value of storage modulus is found to be zero at all frequencies (not shown), whereas loss modulus increases linearly with frequency with a slope of $1.08 \approx 1$, as expected for a pure Newtonian fluid (Figure 2(b)).

**Optical tweezer**: Figure 3(a) shows the measured PSD data for pure water at a laser power of 48 mW. The corner frequency is found to be 27 Hz, giving a trap stiffness 4.64 pN/nm. The corresponding values of loss and storage moduli are calculated using the relation (6) and are shown in Figure 3(b). The frequency dependent loss modulus shows a slope of 1 as expected for a pure viscous liquid. The storage modulus shows a laser power-dependent non zero value as an effect of the trap stiffness up to the corner frequency, after which it reduces to zero, as expected from theory.

The loss and storage moduli of water are also measured using two point microrheology techniques. The value of correlated power spectral density and the resulting loss and storage moduli are shown in Figure 4. We have obtained results similar to that of one point microrheology technique.

**RESULTS AND DISCUSSION**

**Tasar RSF**

The rheological properties of Tasar RSF at two weight percents (w/v) 0.16 and 0.25 are measured by both video microscopy and optical tweezer based techniques (1-point), after applying the corrections explained in the previous sections. The moduli obtained from both the techniques are superimposed on the same plot for each concentration of Tasar RSF (Figure 5).

The video microscopy (optical tweezer) results span the low (high) frequency range from 0.1 to 10 Hz (10 to100Hz), for both loss and storage moduli. At 0.25 weight percent one can see the slight increase of the storage moduli compared to the loss moduli.

The loss moduli and storage moduli of Tasar RSF are also determined from the 2-point microrheological technique as shown in Figure 6. The values obtained from 2-point microrheological technique are comparable with that of 1-point microrheological technique at all frequencies. As our measurement samples are at lower concentration and at these concentrations heterogeneities present in the sample are of long range compared to that of our measurement length scales it is reasonable to expect similar corresponding values using the two techniques.

**Studies on non mulberry silk varieties of Assam**

**Video microscopy**

We have measured the mean squared displacement (MSD) of the dispersed beads in the RSF solution of Eri and Muga. Figure 7(a) shows the MSD of the beads at a concentration 1.00% (w/v). It is observed in the figure that the slope of the MSD lies between 0 and 1, a characteristic of the viscoelasticity of RSF. At lower lag times all curves show slope of nearly 0, a signature of solid-like behavior and at higher lag times, a slope of nearly 1, characteristic of liquid like behavior. Moreover, the probe bead has lower MSD in Muga as compared to Eri, at all lag times. This lower value of the MSD in Muga RSF is a result of the hindered Brownian fluctuations of the beads, compared to that in Eri RSF. The shear modulus is also calculated for the varieties and is shown in Fig. 7(b). The shear modulus of Muga is greater at all frequencies, than that of Eri at the measured concentration of 1.00% (w/v) RSF.

**WAXS measurements**

The XRD diffractograms of the silk fibroin samples were recorded using a X'Pert Pro X-ray diffractometer with Ni filtered, $CuK_\alpha$ radiation of wavelength $\lambda$ = 1.5406 Å, with a graphite monochromator. The scattered beam was focused on a detector. The samples were scanned in the $2\theta$ range 12º-100º with a scanning step size of 0.017º. The profile analysis was carried out using a programme developed by us that employs a multi-dimensional algorithm 'SIMPLEX'. The details of the modeling are given elsewhere [20]. We have obtained the crystal imperfection parameters for exponential distribution functions for each of the silk samples [23].

A graphical part of the crystallite shape ellipse was obtained by taking the crystal size value corresponding to the $2\theta$ = 16.64° direction along the x-axis and the other parameter corresponding to the $2\theta$ = 20.23° direction along the y-axis. The crystallite shape ellipse for the two silk varieties is shown in Fig. 8. The strength of the silk fibre is normally proportional to crystalline area, which is in turn equal to the ellipse area as determined by the micro-structural parameters. It is evident that the crystallite shape ellipse area in Muga is higher than Eri, suggesting that fibres of Muga are stiffer.
The fibre stiffness characterized through crystallite shape ellipses computed from WAXS data reveals an interesting correlation with the corresponding shear moduli of the silk varieties in RSF solution form at the chosen concentration.

## CONCLUSIONS

Our setup is calibrated for both video microscopy and optical tweezer based microrheological measurements. We have measured the viscoelasticity of RSF solution of Tasar silk. Further, we have compared the relative stiffness of Muga and Eri silk fibres obtained by WAXS with the frequency dependent shear moduli of the corresponding RSF solutions.. We observe a clear correlation between the fibre strength and the corresponding shear moduli of the fibres in RSF solution form at 1.00% (w/v).

This interesting correlation may be due to the possibility that the entangled network of the fibre with greater stiffness, offers greater resistance to the tracer bead's movement when it is embedded within a mesh formed by the entangled structures. This results in measuring a larger shear modulus. In future work, we will extend the measurement to a range of concentrations to examine the nature of correlation and to see how it may change.

## ACKNOWLEDGEMENT

We acknowledge a project grant from the Department of Science and Technology, Government of India (Nano mission programme) that enabled this work.

**REFERENCES**

1. Ying, Xu; Yaopeng, Zhang; Huilishao; Xue, Chao, Hu. Int J Biol Macromolecules 2005, 35, 155-161.
2. Rajkhowa, R.; Gupta, V. B.; Kothari, V. K. J Appl Polym Sci 2000, 77(11), 2418.
3. Kushal, Sen; Murugesh Babu, K. J Appl Polym Sci 2004, 92(2) 1080.
4. Kothari, V. K.; Rajkhowa, R.; Gupta, V. B. J Appl Polym Sci 2001, 82(5), 1147.
5. Raghu, A.; Somashekar, R.; Sharath, Ananthamurthy. J Polym Sci Part B: Polym Phys 2007, 45, 2555-2562.
6. Crocker, J.C.; Valentine, M.T.; Weeks, E.R.; Gisler, T.; Kaplan, P.D.; Yodh, A.G.; Weitz, D.A. Phys Rev Lett 2000, 85, 888.
7. Atakhorami, M.; Schmidt, C. F.; Rheol Acta 2006, 45, 449-456.
8. Starrs, L.; Bartlett, P.; Faraday Discuss 2003, 123, 323-334.
9. Xin C.; Knight, D. P.; Shao, Z.; Vollrath, F. Polymer 2001, 42, 9969-9974.
10. Sohn, S.; Strey, H.H.; Gido, S.P, Biomacromolecules 2004, 5, 751-757.
11. Yogesha.; Raghu, A.; Nagesh, B. V.; Sarbari, Bhattacharya.; Mohana, D. C.; Sharath, Ananthamurthy. Int. J. Nanoscience, 2011, (accepted for publication).
12. Atakhorrami, M.; Sulkowska, J. I.; Addas, K. M.; Koenderink, G. H.; Tang, J. X.; Levine, A. J.; MacKintosh, C.; Schmidt, C. F. Phys Rev E 2006, 73, 061501.
13. Gittes, F.; Schnurr, B.; Olmsted, F.; MacKintosh, C.; Schimdt, C. F. Phys Rev Lett 1997, 79, 3286.
14. Crocker, J. C.; Grier, D. G. J. Colloid Interface Sci 1996, 179, 298-310.
15. Apgar, J.; Tseng, Y.; Federov, E.; Herwig, M.B.; Almo, S.C.; Wirtz, D. Biophys J 2000, 79, 1095–1106.
16. IDLVM Code is available on net for free down load http://titan.iwu.edu/~gspaldin/rytrack.html.
17. Waigh, T.A. Rep Prog Phys 2005, 68, 685-742.
18. Mason, T.G.; Weitz, D. A. Phys Rev Lett 1995, 74(7), 1250-1253.
19. Savin, T.; Doyle, P. S.; BioPhys J., 2005, 88, 623.
20. Divakar, S.; Somashekar, R.; Raghu, A.; Yogesha; Sharath, Ananthamurthy; Subrato, Roy. Ind J F Tex Res 2009, 34, 168.
21. Mason, T.G.; Rheol Acta 2000, 39(4), 371-378.
22. Dasgupta, B.R.; Tee, S.Y.; Crocker, J.C.; Frisken, B.J.; Weitz, D. A. Phys Rev E, 2002, 65(5), 051505.
23. Numerical Recipes, edited by Press W, Flannery B P, Teukolsky S & Vetterling W T, (Cambridge University press) 1986.

# FIGURES

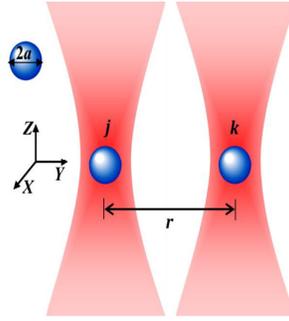

Figure 1: In 2-point microrheology, two beads $j$ & $k$, each of radii '$a$' and separated by a distance of '$r$' are trapped by two laser beams propagating along the '$z$' direction. Beads movement in the '$xy$' plane are tracked by the laser beams backscattered by the two beads, using two independent quadrant detectors.

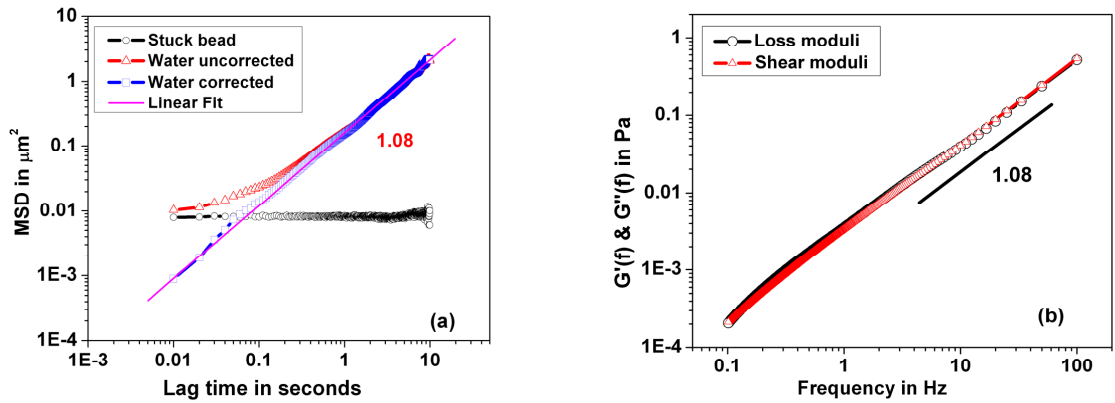

Figure 2: (a) Ensemble average of MSD of beads in water as a function of lag time, before (triangles) and after (squares) subtraction of static error. Corrected MSD of beads in water against lag time shows a slope of 1.08. MSD of beads stuck to the cover slip is also shown in the plot. (b) Frequency dependent shear and loss moduli of water estimated from the corrected data is shown in the plot (a).

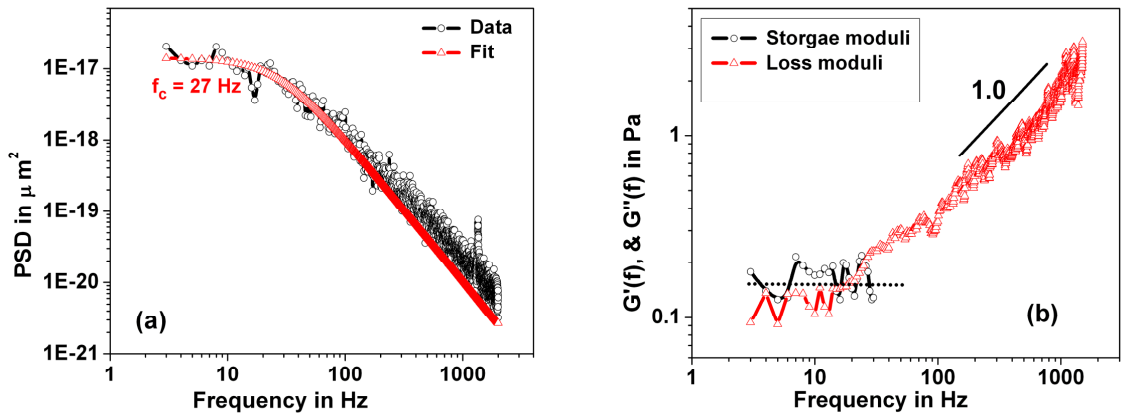

Figure 3: a) Power spectral density of a trapped bead in water at a laser power of 48 mW at the sample plane. b) Frequency dependent loss and storage moduli of water by optical tweezer based technique; the plateau in the graph of the storage moduli at lower frequencies is due to the stiffness of the trap, which is found to be 0.146 Pa at 48 mW laser power (shown in dotted line).

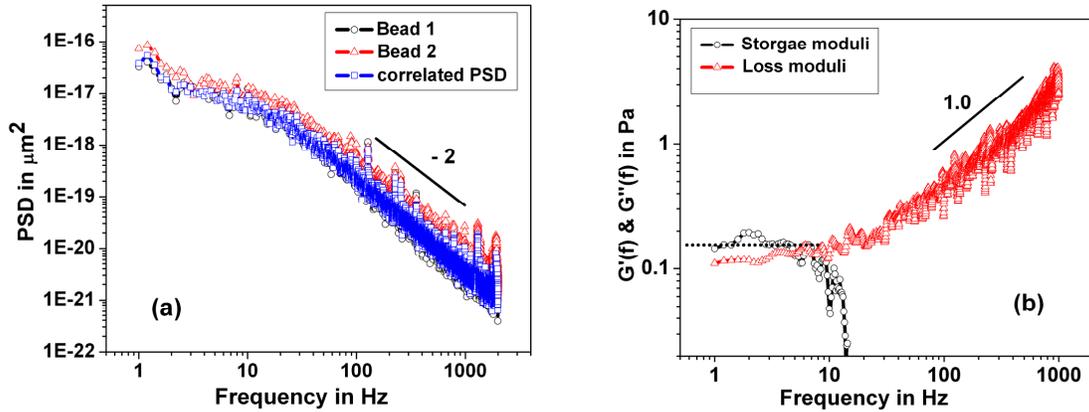

Figure 4: (a) PSD data obtained from the correlated motion. (b) Corresponding frequency dependent loss and storage moduli of water by optical tweezer based technique. The finite storage modulus measured is because of the trap stiffness and this drops to zero at higher frequencies.

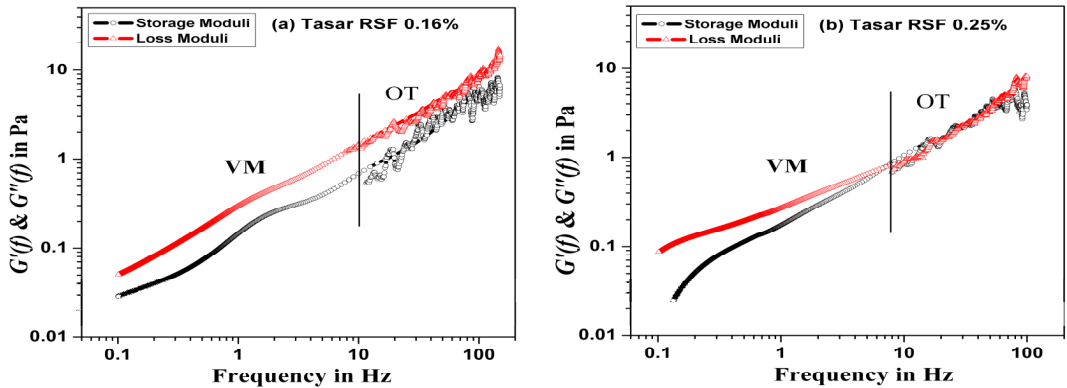

Figure 5: Frequency dependent loss and storage moduli of Tasar RSF at (a) 0.16 and (b) 0.25 weight percents from video microscopy and optical tweezer techniques.

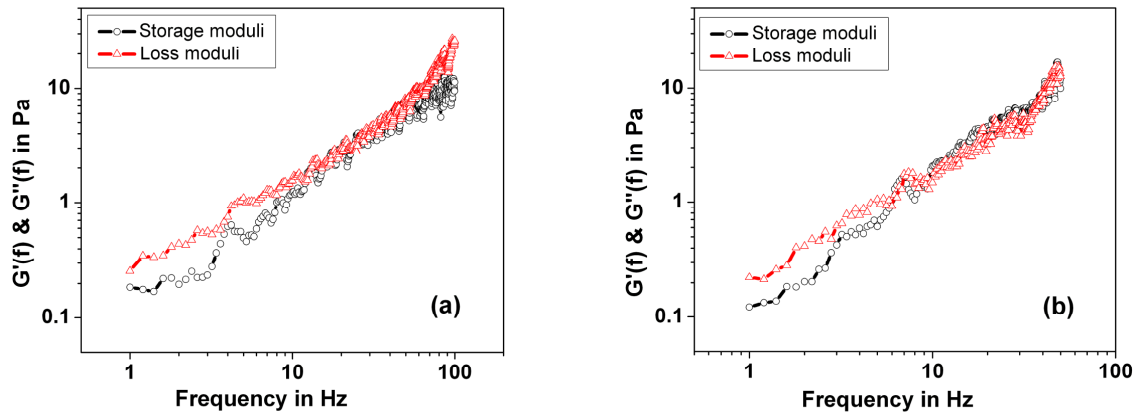

Figure 6: Frequency dependent loss and storage moduli of Tasar RSF at (a) 0.16 and (b) 0.25 weight percent from optical tweezer based 2 point microrheology technique.

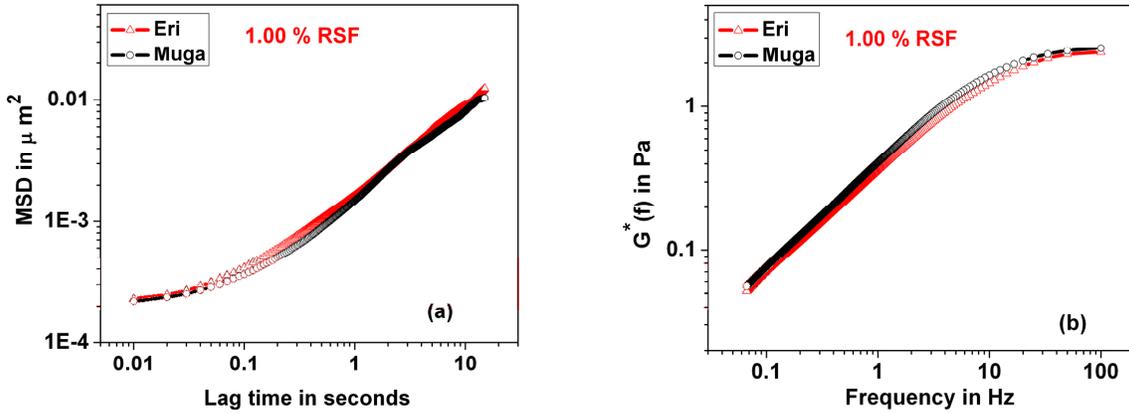

Figure 7: (a) Ensemble average of MSD of beads in RSF of Muga and Eri. (b) Corresponding Shear moduli for three samples at a concentration of 1.00 %( w/v). The errors on these values are very small and are not shown as they are within the symbol size denoting the data point.

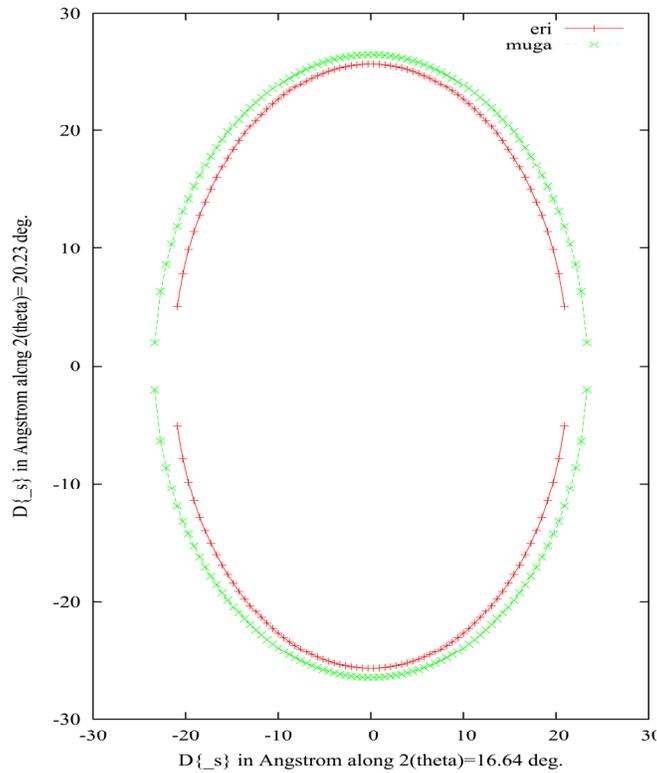

Figure 8: Variation in crystallite shape ellipse for non mulberry silk fibroins viz., Muga and Eri.